# Complex cooperativity of ATP hydrolysis in the $F_1$-ATPase molecular motor

Ming S. Liu, B. D. Todd and Richard J. Sadus

*Centre for Molecular Simulation, Swinburne University of Technology,
P.O.Box 218, Hawthorn, Victoria 3122, Australia*

**Abstract:**

$F_1$-ATPase catalyses ATP hydrolysis and converts the cellular chemical energy into mechanical rotation. The hydrolysis reaction in $F_1$-ATPase does not follow the widely believed Michaelis-Menten mechanism. Instead, the hydrolysis mechanism behaves in an ATP-dependent manner. We develop a model for enzyme kinetics and hydrolysis cooperativity of $F_1$-ATPase which involves the binding-state changes to the coupling catalytic reactions. The quantitative analysis and modeling suggest the existence of complex cooperative hydrolysis between three different catalysis sites of $F_1$-ATPase. This complexity may be taken into account to resolve the arguments on the binding-change mechanism in $F_1$-ATPase.

# 1. Introduction

$F_1F_0$-ATPase is an enzyme complex that is vital to cellular energy conversion. It works as a dual-domain molecular motor powered by two types of driving forces [1-4]. The off-membrane $F_1$ domain synthesizes adenosine triphosphate (ATP) from adenosine diphosphate (ADP) and inorganic phosphate (Pi) driven by a proton-motive force from the in-membrane $F_0$ domain. Alternatively, when operating in the reverse, it hydrolyzes ATP into ADP and Pi at $F_1$ and releases energy. In both cases, coupling of the transduction of chemical energy from catalysis reactions with the conformational changes generates rotary mechanical torque at $F_1$. Isolated $F_1$-ATPase can also work independently as a rotating motor fueled by ATP hydrolysis [4]. Cloned $F_1$-ATPases have been explored as biologically powered nano-machines because they work at an exceptionally high chemomechanical coefficient [5-9].

The functionally stable $F_1$-ATPase is composed of subunits designated as $\alpha$, $\beta$, $\gamma$, $\delta$, and $\varepsilon$ with a stoichiometry of 3:3:1:1:1 [10-12]. The $\alpha$ and $\beta$ subunits alternate in an hexagonal arrangement around a central cavity containing the N- or C-terminal helices of $\gamma$, $\delta$ and $\varepsilon$ subunits, as illustrated in Figure 1. Three alternative sites on the hexamer formed by subunits $(\alpha\beta)_3$ are found to be catalytically active and responsible for ATP hydrolysis/synthesis [1,2,10,11]. These catalytic sites are located on the $\beta$ subunits at the interfaces with the $\alpha$ subunits. The crystal structures of three $(\alpha\beta)$ pairs are almost identical and have a strong symmetry, but the incorporation of the central $\gamma\delta\varepsilon$ subunits creates a structural and functional asymmetry between the three catalytic sites [1,2,4,10,11].

Long before the molecular level structures and binding-states were determined, Boyer proposed a 'bi-site' binding change mechanism [13] for rapid hydrolysis of $F_1F_0$-ATPase based on its kinetics. As shown in Figure 1(a), three $(\alpha\beta)_3$ catalysis sites are in different conformations at any given time, namely they bind with ATP, or ADP and/or Pi, or they are empty. The existence of alternative conformations depends on their respective positions relative to the concave, neutral, or convex sides of the central shaft $\gamma$ subunit. The catalytic sites at three $(\alpha\beta)$ pairs work in a sequential collaboration – while one site tightly binds ATP and undergoes hydrolysis, the next one loosely binds the hydrolyzed

ADP, and the third one opens to release the hydrolysis products and intake ATP. The collaborative conformational changes in $(\alpha\beta)_3$ induce a torque between the hexamer $(\alpha\beta)_3$ and the central stalk $\gamma$ subunit, causing the $F_1$-ATPase motor to rotate [1-2]. Atomistic structural investigations [10,11], along with motor experiments of $F_1$-ATPase (especially the visualized rotation by Noji et al. [5,6]), confirmed Boyer's mechanism. Subsequently other issues were raised concerning the multisite binding-change scheme and the cooperativity of hydrolysis reactions at multiple catalytic sites [12,14,15,30,31]. For example, the issue of whether or not a bi-site mechanism is universally valid remains unresolved [12,14]. Contrary to the belief of many researchers, the kinetics of a rotating $F_1$-ATPase motor does not follow Michaelis-Menten kinetics [6,7]. Considerable efforts have been reported to resolve these issues and to reveal the rotating chemomechanic nature of $F_1$-ATPase. These works include sophisticated single molecule measurements [6,7,9,16-18], chemomechanic modeling [19-23] and atomistic simulation using molecular dynamics methods [24-27]. However, the dynamic and interactive cooperativity at different ATP catalysis sites has not been convincingly demonstrated. The molecular coupling mechanism of ATP hydrolysis in $F_1$-ATPase is still beyond fully understanding.

Previously, we proposed a model of the $F_1$-ATPase motor based on enzyme kinetics and rotary Langevin dynamics [23]. In the model, the energy transduction and stepwise rotation of $F_1$-ATPase were regulated by a series of near-equilibrium reactions when nucleotides bind or unbind. For the case of $F_1$-ATPase driving an actin filament, the theoretical load-rotation profiles were analyzed against experiment and gave good agreement. The link from the molecular-scale hydrolysis reactions to the micro-scale mechanical rotation of $F_1$-ATPase was established without considering the possible effect from the non-Boyer schemes. In this work, we quantitatively analyze the complex cooperativity of ATP hydrolysis and describe non-Boyer chemomechanics of the $F_1$-ATPase molecular motor.

## 2. Kinetic model of hydrolysis cooperativity in $F_1$-ATPase

The binding conformations of the multiple catalysis sites of $F_1$-ATPase are identified as $\alpha\beta_{TB}$, $\alpha\beta_{LB}$, and $\alpha\beta_O$ [1,2, 10-12] (Figure 1). The open (O) site has a very low affinity for substrates of ATP, ADP or Pi and is catalytically inactive, whereas the other sites involve either a loosely bound (LB) substrate ADP or a tightly bound (TB) substrate ATP. The actual order of unbinding and release of ADP and Pi were proved to be Pi first, followed by ADP [4,17]. Therefore a complete enzymatic cycle of ATP hydrolysis in $F_1$-ATPase occurs via the following pathway,

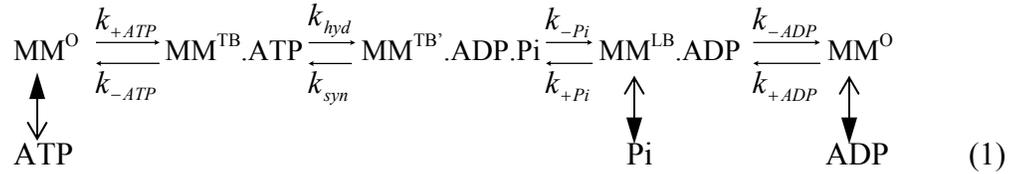

(1)

where $k_{+ATP}$, $k_{-ATP}$ (or, $k_{+ADP}$, $k_{-ADP}$ and $k_{+Pi}$, $k_{-Pi}$) refer to the rate constants of association and dissociation of ATP (or, ADP and Pi) molecules to the motor, i.e., they are the rate of ATP (or, ADP and Pi) binding to or unbinding from enzymatic-active sites of $(\alpha\beta)_3$. $k_{hyd}$, $k_{syn}$ are the hydrolysis and synthesis rate constants of ATP. In this catalysis pathway, the direction of the hydrolysis resulting in binding/unbinding to the motor of ATP, ADP and Pi is given by the solid arrows, whereas the thin arrows point to the synthesis direction.

Based on Boyer's mechanism, the binding changes of $F_1$-ATPase can be highly coupled to an ordered ATP hydrolysis at three catalysis sites [23]. In each hydrolysis cycle, the three ($\alpha\beta$) sites interact in a cooperative way as the central stalk $\gamma$ sequentially rotates relative to them. In steady cycles of ATP hydrolysis reactions, three ($\alpha\beta$) pairs cooperate with respect to their conformational changes in a synchronized way. For the enzyme kinetics of $F_1$-ATPase, the multi-site hydrolysis at three ($\alpha\beta$) catalysis subunits are believed to be biochemically equal [1,2,13]. The hydrolysis reactions at three sites could be regarded as a three-fold fast near-equilibrium process with the same reaction sequence and dynamics. Therefore we interpret the steady state ATP hydrolysis cycles of Eq.(1) by a series of fast equilibrium reactions,

$$\begin{cases} P_O + P_{ATP} + P_{ADP.Pi} + P_{ADP} = 1 \\ k_{ATP} \cdot [ATP] \cdot P_O + k_{syn} \cdot P_{ADP.Pi} - (k_{hyd} + k_{-ATP}) \cdot P_{ATP} = 0 \\ k_{hyd} \cdot P_{ATP} + k_{Pi} \cdot [Pi] \cdot P_{ADP} - (k_{syn} + k_{-Pi}) \cdot P_{ADP.Pi} = 0 \\ k_{-Pi} \cdot P_{ADP.Pi} + k_{ADP} \cdot [ADP] \cdot P_{open} - (k_{Pi} \cdot [Pi] + k_{-ADP}) \cdot P_{ADP} = 0 \\ k_{-ATP} \cdot P_{ATP} + k_{-ADP} \cdot P_{ADP} - (k_{ATP} \cdot [ATP] + k_{ADP} \cdot [ADP]) \cdot P_O = 0 \end{cases} \quad (2)$$

Here [ATP], [ADP] and [Pi] are the concentrations of ATP, ADP and Pi respectively. $P_O$, $P_{ATP}$, $P_{ADP.Pi}$ and $P_{ADP}$ are the probability of the states when different pairs of $(\alpha\beta)_3$ are either empty or occupied by ATP, ADP.Pi or ADP molecules, respectively (as illustrated in Figure 1). Here we assume that the $F_1$-ATPase motor is under physiological conditions and ATP, ADP and Pi molecules are fully dissolved in solution. Eq. (2) is the steady-state cycle equivalent of Eq. (1), which is justified by the fact that the alternating binding states and the chemical reaction steps are repeatedly coupled. It should be noted that metal ions, in particular $Mg^{2+}$, have a crucial role in ATPase enzymatic activity for a physiological catalysis. The different concentration of $Mg^{2+}$ does affect the kinetics of $F_1$-ATPase by coordinating ATP catalysis [37]. However, since $Mg^{2+}$ changes the rate constants but does not alter the catalysis pathways, the present kinetic model naturally includes any possible effect of $Mg^{2+}$ via the rate constants.

The steady-state condition of Eq. (1) gives the overall reaction rate of ATP hydrolysis at given physiological conditions (such as, of given [ATP], [ADP] and [Pi]) by,

$$R = k_{hyd} P_{ATP} - k_{syn} P_{ADP \cdot Pi} \tag{3a}$$

For $F_1$-ATPase following Boyer's mechanism, this leads to [23],

$$R = \left(-k_{-ATP} k_{syn} k_{ADP}[ADP] k_{Pi}[Pi] + k_{-ADP} k_{-Pi} k_{hyd} k_{ATP}[ATP]\right) \Big/$$
$$\left(k_{-ADP} k_{-Pi} k_{hyd} + (k_{hyd} k_{-Pi} + k_{-ATP} k_{-Pi} + k_{-ATP} k_{syn} + (k_{-ATP} + k_{syn} + k_{ADP}) k_{Pi}[Pi]) k_{ADP}[ADP] + \right.$$
$$\left. k_{hyd}(k_{-ADP} + k_{-Pi} + k_{Pi}[Pi]) k_{ATP}[ATP] + (k_{-ADP} k_{-Pi} + k_{-ADP} k_{syn} + k_{syn} k_{Pi}[Pi])(k_{-ATP} + k_{ATP}[ATP])\right)$$
$$\tag{3b}$$

However, the symmetric $(\alpha\beta)_3$ structures and nearly identical catalysis kinetics can not guarantee that the catalysis at three sites takes place in a highly ordered fashion. To check the possibility of complex cooperativity between three catalytic sites of $F_1$-ATPase, we may carefully investigate the cooperativity from its enzyme kinetics or motor chemomechanics. From a kinetics point of view, the enzymatic cooperativity may be quantitatively measured by a modified Hill Equation [28],

$$\log(R/(R_{max} - R)) = h \log[ATP] - \log K \tag{4}$$

where $R_{max}$ is the saturation rate (or motor speed if applicable) in $F_1$-ATPase, $K$ is a constant and $h$ is the Hill number. When $h = 1$, there is no cooperative interaction between different catalytic sites; only one ATP molecule is hydrolyzed at any time and the kinetics follows the Michaelis-Menten mechanism, i.e. $R = R_{max}[ATP]/(K_m +[ATP])$ with $K_m$ the Michaelis constant. When $h > 1$, there is positive cooperative hydrolysis between different sites and more than one ATP molecule is simultaneously undergoing catalysis reactions. Values of $h < 1$ indicate negative cooperative behavior where there might be some inhibition between the multi catalysis sites.

The calculations involving Eq. (2) to Eq. (4) were programmed using the Mathematica library [36] and executed on a standard pentium-4 PC or Unix workstation. The overall computing time is of the order of CPU minutes. This represents a considerable advantage over time-costly atomistic simulation [24-27].

## 3. Results and discussions

In Boyer's 'bi-site' activation (Figure 1(a)), three catalysis sites are in different reaction stages (binding states), and on average only one catalytic site carries out a specific hydrolysis reaction step at any one time. Therefore in Hill's plot, bi-site catalysis should give $h = 1$. There is currently little evidence that the chemical reaction steps in ATP synthase take place on average at three different states at a time. On the contrary, solid evidence from chloroplast [29], *E. Coli* [30,31], and bovine mitochondrial $F_1$-ATPase [12] suggests that two or more sites may be active simultaneously, as schematized in Figure 1(b). For example, in chloroplast $F_1$-ATPase [29], binding states of an ADP analog to cooperating sites depends on nucleotide concentration. Due to very strong cooperativity, two of the three sites were found simultaneously to be in the tightly bound state. In addition, both the two $K_m$ constants fit to the reaction rate [6,7] and our simulation (Figure 4 of ref. [23]) of the mechanochemistry of *Bacillus PS3* $F_1$-ATPase implied that more than one ATP molecule are needed at a time for the steady operation of the $F_1$-ATPase motor. These considerations strongly suggest that a non-Boyer mechanism does occur in $F_1$-ATPase.

To determine the difference between Boyer-like and non-Boyer-like cooperative hydrolysis in $F_1$-ATPase, we firstly re-analyze the motor dynamics of *Bacillus PS3* $F_1$-ATPase [6,7] for possible enzymatic cooperativity schemes. The kinetic simulations from both Boyer-like and non-Boyer-like mechanisms are performed to interpret the complex cooperativity. We also calculate the occupation probability of different binding states. This calculation may help to determine the bi-site or tri-site catalysis mechanism for $F_1$-ATPase operating in different [ATP] regions.

3.1 Complex cooperativity of ATP hydrolysis for multiple catalysis sites of $F_1$-ATPase

Figure 2 shows the Hill plots, $log(R/(R_{max}-R))$ versus $log$[ATP], of ATP hydrolysis rate for the $F_1$-ATPase motor. The experimental results (diamonds [6,7]) are re-produced in the Hill manner. Both the experiments and our simulations of strong cooperativity (solid line) indicate that ATP hydrolysis in $F_1$-ATPase does not simply follow the Michaelis-Menten mechanism, which should be parallel to the dot-dashed line (i.e., $h = 1$). There is different cooperativity of ATP hydrolysis from multiple catalysis sites occurring across the whole [ATP] region.

From the Hill plot (Figure 2) of the experimental data, the cooperativity of hydrolysis reactions in $F_1$-ATPase might be classified into three concentration regions. Region 1, for the lower region ([ATP] = nM ~ 10 μM), $F_1$ obeys Michaelis-Menten kinetics and there are no cooperative hydrolysis reactions. Region 2 is in the intermediate pre-saturated region ([ATP] = 10 μM ~ 100 μM), and there is a negative cooperativity. Region 3 is in the saturated region ([ATP] > 100 μM), and there is a strong cooperativity between the hydrolysis reactions from different sites. In Region 1, a Boyer scheme simulation (the crossed line, with a full set of rate constants from Panke et al. [21]) is sufficient if we take the binding rate of ATP as determined by experiment [6,7]. However, from the pre-saturated Region 2 to the saturated Region 3, a complex cooperativity of ATP hydrolysis is apparent.

To perform the simulation, we applied the hypothesis that the number of multiple catalysis sites is unknown except that their pathway of hydrolysis reactions is known as Eq. (1). These catalytic sites are also assumed to have the same rate of association and dissociation of nucleotides of ATP, ADP.Pi or ADP. When three ATPs are hydrolyzed simultaneously in the same pathway at three catalysis sites, the solid line in Figure 2 gives a much better fit to the experimental data (in Region 3). This indicates that strong cooperativity of ATP hydrolysis is the case for the $F_1$-ATPase motor, at least for the saturated [ATP] region. Region 2 seems to have a transition from bi-site to tri-site catalysis. A chemomechanical and molecular structural mechanism for this transition remains unknown.

For comparison, Figure 3 illustrates the simulations of the overall hydrolysis reaction rates of $F_1$-ATPase versus [ATP] for different schemes of hydrolysis. For Region 3 ([ATP] > 100 μM), the simulation of ATP hydrolysis with strong 'tri-site' cooperativity (solid lines in Figures 2 and 3) agrees much better with experimental data for the *Bacillus PS3* $F_1$-ATPase motor (diamonds [6,7]), compared to the bi-site catalysis simulations (dash and crossed lines in Figures 2 and 3). It is interesting to note that complex cooperativity of ATP hydrolysis does not only exist in $F_1$-ATPase, but also occurs in other ATP-fueled multi-domain molecular motors, such as myosin (Liu, M.S., Todd, B.D. and Sadus, R.J., unpublished data). This appears to be a common chemomechanical character for the ATP-fueled molecular motors.

3.2 Bi-site vs tri-site hydrolysis for $F_1$-ATPase

Strong cooperativity of ATP hydrolysis in $F_1$-ATPase will require 'tri-site' catalysis, such as depicted by a possible scheme in Figure 1(b). In a tri-site mechanism, ATP hydrolysis probably plays a dominant role over ATP binding in producing rotary torque, and the bound nucleotide persists on the enzyme through more than one turnover for a subunit rotation step. An additional fifth conformation might have to be considered beyond the conformations in Eq. (1). This would most likely be a loosely-bounded site with enhanced affinity for Pi plus ADP and concomitantly lowered affinity for tightly bound ATP or ADP.Pi. The net result would favor Pi + ADP binding rather than ADP binding (to an empty site). This new scheme might help to explain the half-open ADP-bound structure [12] and the sub-steps of rotation of the $F_1$-ATPase motor [7]. Nevertheless, such a mechanism may lead to incompatible chemomechanics with some motor experiments [32], in which the same 'bi-site' mechanism is believed to prevail throughout the whole [ATP] region [6,7]. Boyer argued [32] that the third site occupation during high [ATP] hydrolysis can be explained by rebinding or retention of ADP. He emphasized that bi-site activation manifests the characteristics of only one catalytic pathway in $F_1$-ATPase [32]. However, even if only one type of catalytic site exists at any one time, it does not exclude the possibility of two or more sites simultaneously carrying out the same reaction pathway. The question is how are the hydrolysis reactions of $F_1$-ATPase coupled to changes from uni-site, bi-site or tri-site binding-states? The answer is currently hindered by the absence of direct determination of the binding states and their changes during steady hydrolysis [14].

It is biochemically straightforward for $F_1$-ATPase to undergo both unisite catalysis and multisite catalysis. When [ATP] is in sub-stoichiometric quantity, it binds to the first site with very high affinity. As this ATP is hydrolyzed to ADP + Pi, the products are released slowly ($k_{off} < 10^{-3}$ s$^{-1}$). The reversible hydrolysis/synthesis reaction occurs with an equilibrium constant close to 1, and $F_1$ undertakes a unisite catalysis. Multisite catalysis occurs when there is high enough [ATP] and ATP begins to bind to the next site. Therefore, we need to re-examine the above question for $F_1$-ATPase, namely, whether the binding is 'bi-site' or 'tri-site', and how do the binding-state changes couple with rapid hydrolysis. For the distribution of occupancy of catalysis sites in *E. Coli* $F_1$-ATPase, Senior and coworkers [14,18,30,31] found that at lower [ATP] (< 1 μM), when primarily the high affinity site was occupied, the equilibrium reaction is a unisite catalysis. At high [ATP] (>>1 μM), all three catalysis sites are occupied and filled with ATP or ADP and tri-site hydrolysis takes place. The probability of an empty site is rare. Thus Senior concluded that a 'bi-site' mechanism does not exist [31].

Our theory permits one to calculate the occupation probability of different binding states during steady ATP hydrolysis. The simulations, as shown in Figure 4, indicate that the binding state distribution changes dramatically as a function of [ATP] before the hydrolysis is saturated. In other words, the hydrolysis mechanism is [ATP] dependent. Figure 4 shows the probability of binding state versus [ATP] for any single catalysis cycle at a given site of chloroplasts $F_1$-ATPase (when three sites follow the same kinetics pathway of Eq. (1)). Even if no cooperativity between three catalysis sites occurs, Boyer's bi-site scheme (where the probability of $P_{open} = P_{ADP} = P_{ATP} + P_{ADP.Pi} = 1/3$, [33]) is only valid for a limited region. It is approximately at [ATP] of 100 μM ~ 200 μM for the chloroplasts $F_1$-ATPase motor, which is the pre-saturated region. For higher [ATP], $F_1$-ATPase undergoes saturated rapid hydrolysis and the probability of a site being empty is small (10% or less at saturated region). In this case, $P_{ADP} \cong 2/3$ and $P_{ATP}+P_{ADP.Pi} \cong 1/3$ and all three catalysis sites are in bounded states. This is consistent with the experimental observation [30,31], and in this concentration region $F_1$-ATPase undertakes tri-site hydrolysis.

## 4. Conclusion

For ATP hydrolysis of the $F_1$-ATPase motor, our study and the comparison with experimental data indicate that there is: 1) a complex cooperativity between three catalytic sites for rapid ATP hydrolysis; 2) different rate constants of association, $k_{+ATP}$, might exist (at least for very low [ATP]); and 3) the binding-change mechanism is [ATP] dependent, changing from uni-site for lower [ATP], to bi-site for pre-saturated [ATP] and to tri-site at high [ATP]. The complexity of cooperative ATP catalysis is at present difficult to be revealed experimentally (mainly limited by the lack of suitable technology), given that ATP reactions at different catalysis sites are affected strongly by dynamic changes of binding affinity and constant fluctuation in a confined space. The complex cooperativity in $F_1$-ATPase is expected to be resolved with more comprehensive atomistic structural investigations and single molecular measurements, such as to probe disordered kinetics and enzymatic dynamics at the single molecular level [35].

*Acknowledgement*:
We thank the Australian and Victorian Partnerships for Advanced Computing for generous allocation of computing resources. Dr Matthew Downton is thanked for helpful discussion.

## References:

[1] P. D. Boyer, The ATP synthase – A splendid molecular machine, Annu. Rev. Biochem. **66** (1997) 717–749.
[2] P.D. Boyer, What makes ATP synthase spin? Nature **402** (1999) 247-249.

[3] W. Junge, ATP synthase and other motor proteins, Proc. Natl. Acad. Sci. USA **96** (1999) 4735-4737.

[4] K. Kinosita, Jr., R. Yasuda, H. Noji, and K. Adachi, A rotary molecular motor that can work at near 100% efficiency, Philos. Trans. R. Soc. Lond. B**355** (2000) 473-489.

[5] H. Noji, R. Yasuda, M. Yoshida, and K. Kinosita, Direct observation of the rotation of $F_1$-ATPase, Nature **386** (1997) 299–302.

[6] R. Yasuda, H. Noji, K. Kinosita, and M. Yoshida, $F_1$-ATPase is a highly efficient molecular motor that rotates with discrete 120˚ steps, Cell **93** (1998) 1117–1124.

[7] R. Yasuda, H. Noji, M. Yoshida, K. Kinosita, and H. Itoh, Resolution of distinct rotational substeps by submillisecond kinetic analysis of $F_1$-ATPase, Nature **410** (2001) 898-904.

[8] R.K. Soong, G.D. Bachand, H.P. Neves, A.G. Olkhovets, H.G. Craighead, and C.D. Montemagno, Powering an inorganic nanodevice with a biomolecular motor, Science **290** (2000) 1555-1558.

[9] O. Pänke, D. A. Cherepanov, K. Gumbiowski, S. Engelbrecht, and W. Junge, Viscoelastic dynamics of actin filaments coupled to rotary F-ATPase: Angular torque profile of the enzyme, Biophys J. **81** (2001) 1220-1233.

[10] J. P. Abrahams, A. G. W. Leslie, R. Lutter, and J. E. Walker, Structure at 2.8Å of $F_1$-ATPase from bovine heart mitochondria, Nature **370** (1994) 621–628.

[11] D. Stock, A.G.W. Leslie, and J.E. Walker, Molecular architecture of the rotary motor in ATP synthase, Science **286** (1999) 1700-1705.

[12] R. Ian Menz, J.E. Walker, and A.G.W. Leslie, Structure of bovine mitochondrial $F_1$-ATPase with nucleotide bound to all three catalytic sites: implications for the mechanism of rotary catalysis, Cell **106** (2001) 331-341.

[13] P.D. Boyer, in: L. Emster, R. Estabrook and E.C. Slater (Eds.) Dynamics of Energy Transducing Membranes, Elsevier, Amsterdam, 1974, pp. 289–301. Also, P.D. Boyer and W.E. Kohlbrennerin, in: B. R. Selman and S. Selman-Reimer (Eds.) Energy Coupling in Photosynthesis, Elsevier, New York, 1981, pp. 231–240.

[14] A.E. Senior, S. Nadanaciva, and J. Weber, The molecular mechanism of ATP synthesis by $F_1F_0$-ATP synthase, Biochim. Biophys. Acta **1553** (2002) 188-211.

[15] H. Noji and M. Yoshida, The rotary machine in the cell, ATP synthase, J. Biol. Chem. **276** (2001) 1665-1668.

[16] Y.M. Milgrom, M.B. Murataliev, and P.D. Boyer, Bi-site activation occurs with the native and nucleotide-depleted mitochondrial $F_1$-ATPase, Biochem. J. **330** (1998) 1037-1043.

[17] T. Masaike, E. Muneyuki, H. Noji, K. Kinosita, Jr., and M. Yoshida, $F_1$-ATPase changes its conformations upon phosphate release, J. Biol. Chem. 277 (2002) 21643–21649.

[18] J. Weber, S. Wilke-Mounts, and A.E. Senior, Quantitative determination of binding affinity of delta-subunit in *Escherichia coli* $F_1$-ATPase: effects of mutation, $Mg^{2+}$, and pH on Kd, J. Biol. Chem. **277** (2002) 18390- 18396.

[19] H. Y. Wang and G. Oster, Energy transduction in the $F_1$ motor of ATP synthase, Nature **396** (1998) 279–282.

[20] H. Y. Wang and G. Oster, The Stokes efficiency for molecular motors and its applications, Europhys. Lett. **57** (2002) 134-140.

[21] O. Pänke and B. Rumberg, Kinetic modeling of rotary $CF_0F_1$-ATP synthase: storage of elastic energy during energy transduction, Biochim. Biophys. Acta **1412** (1999) 118-128.

[22] D.A. Cherepanov and W. Junge, Viscoelastic dynamics of actin filaments coupled to rotary F-ATPase: Curvature as an indicator of the torque, Biophys J. **81** (2001) 1234–1244.

[23] M.S. Liu, B.D. Todd, and R.J. Sadus, Kinetics and chemomechanical properties of the $F_1$-ATPase molecular motor, J. Chem. Phys. **118** (2003) 9890-9898.

[24] R.A. Bockmann and H. Grubmuller, Nanoseconds molecular dynamics simulation of primary mechanical energy transfer steps in $F_1$-ATPase, Nature Struct. Biol. **9** (2002) 198-202.

[25] J. P. Ma, T.C. Flynn, Q. Cui, A.G.W. Leslie, J.E. Walker and M. Karplus, A dynamic analysis of the rotation mechanism for conformational change in $F_1$-ATPase, Structure **10** (2002) 921-931.

[26] I. Antes, D. Chandler, H. Wang, and G. Oster, The unbinding of ATP from F1-ATPase, Biophys. J. **85** (2003) 695-706.

[27] M. Dittrich, S. Hayashi and K. Schulten, On the mechanism of ATP hydrolysis in $F_1$-ATPase, Biophys. J. **85** (2003) 2253-2266.

[28] A. Fersht, Structure and Mechanism in Protein Science: A Guide to Enzyme Catalysis and Protein Folding, W.H. Freeman & Co., New York, 1999, Chapter 10.

[29] S. Gunther and B. Huchzermeyer, Nucleotide binding of an ADP analog to cooperating sites of chloroplast $F_1$-ATPase ($CF_1$), Eur. J. Biochem **267** (2000) 541-548.

[30] J. Weber and A.E. Senior, Catalytic mechanism of $F_1$-ATPase, Biochim. Biophys. Acta **1219** (1997) 19-58.

[31] J. Weber and A.E. Senior, Bi-site catalysis in $F_1$-ATPase: does it exist? J. Biol. Chem. **276** (2001) 35422- 35428.

[32] P.D. Boyer, Catalytic site occupancy during ATP synthase catalysis, FEBS Lett. **512** (2002) 29-32.

[33] In $F_1$-ATPase, ATP hydrolysis and reversible synthesis are at fast equilibrium. $O_{18}$ exchange studies confirmed that ATP and ADP.Pi interconvert rapidly in the active site [1,13]. To date the number of nucleotides and, in particular, the conformations bounded with ATP or ADP.Pi are not dynamically distinguishable [14, 34], unless Pi or ADP is further released from the active site.

[34] K. Dong, H. Ren, and W. S. Allison, The fluorescence spectrum of the introduced tryptophans in the $\alpha_3(\beta F155W)_3\gamma$ subcomplex of the $F_1$-ATPase from the thermophilic *Bacillus* PS3 cannot be used to distinguish between the number of nucleoside di- and triphosphates bound to catalytic sites, J. Biol. Chem. **277** (2002) 9540-9547.

[35] X. Sunney Xie, Single-molecule approach to dispersed kinetics and dynamic disorder: Probing conformational fluctuation and enzymatic dynamics, J. Chem. Phys. **117** (2002) 11024-11032.

[36] Mathematica ver5.0 is a product of Wolfram Research, Inc (www.wolfram.com).

[37] J. Weber, S.T. Hammond, S. Wilke-Mounts, and A.E. Senior, $Mg^{2+}$ coordination in catalytic sites of $F_1$-ATPase, Biochemistry **37** (1998) 608-614.

Figure captions:

Figure 1. Conceptual schemes of the time-averaged binding states occupation during steady hydrolysis of $F_1$-ATPase. (a) the 'bi-site' activation proposed by Boyer, and (b) one possible scheme of non-Boyer 'tri-site' activations. $(\alpha\beta_T)$ $(\alpha\beta_L)$ and $(\alpha\beta_O)$ refer to the conformational tight-binding, loose-binding, and open states of $(\alpha\beta)$ pairs, respectively.

Figure 2. The Hill plots of ATP hydrolysis rate, $log(R/(R_{max}-R))$, versus $log$[ATP] show non-Michaelis-Menten kinetics in $F_1$-ATPase. The diamonds refer to the experimental results [6,7]. The dashed and crossed lines are simulations according to Boyer's bi-site scheme. The solid line indicates simulation when strong cooperativity of ATP hydrolysis occurs (three ATPs are hydrolyzed simultaneously). The rate constants are the same as in Figure 3.

Figure 3. Overall hydrolysis reaction rate, $R$, of $F_1$-ATPase versus ATP concentration. The dashed line is the simulated hydrolysis rate according to Boyer's bi-site scheme [23] with $k_{ATP} = 2.08\times10^6$ $M^{-1}s^{-1}$, $k_{ADP} = 8.90\times10^6$ $M^{-1}s^{-1}$, $k_{Pi} = 8.10\times10^5$ $M^{-1}s^{-1}$, $k_{-ATP} = 2.70\times10^2$ $s^{-1}$, $k_{-ADP} = 4.90\times10^2$ $s^{-1}$, $k_{-Pi} = 2.03\times10^3$ $s^{-1}$, and $k_{hyd} = 4.5\times10^5$ $s^{-1}$, $k_{syn} = 1.15 \times 10^{-3}$ $s^{-1}$ (from Panke et al. [21]). The crossed line is again the bi-site simulation with a determined association rate [7] of ATP, $k_{ATP} = 3.0\times10^7$ $M^{-1}s^{-1}$. The solid line is the simulation for the case of strong cooperativity between three catalysis sites, and the diamonds refer to experimental data [6,7]. ([ADP] and [Pi] are set at cellular physiological condition of 10 μM and 1.0 mM, respectively).

Figure 4. Occupation probability of binding-state changes versus [ATP]. ATP-dependent binding states imply complex cooperativity of ATP hydrolysis for three different catalysis sites. Boyer's bi-site catalysis scheme only holds for a limited [ATP] region, around 150 μM ~ 250 μM in this case. The rate constants are the same as in Figure 3 ([ADP] and [Pi] are consistently set as 10 μM and 1.0 mM, respectively).

(a) 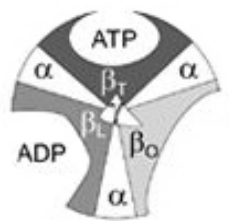

(b) 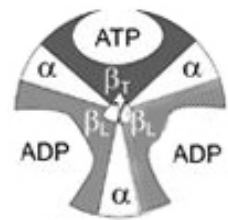

Liu et al., Figure 2

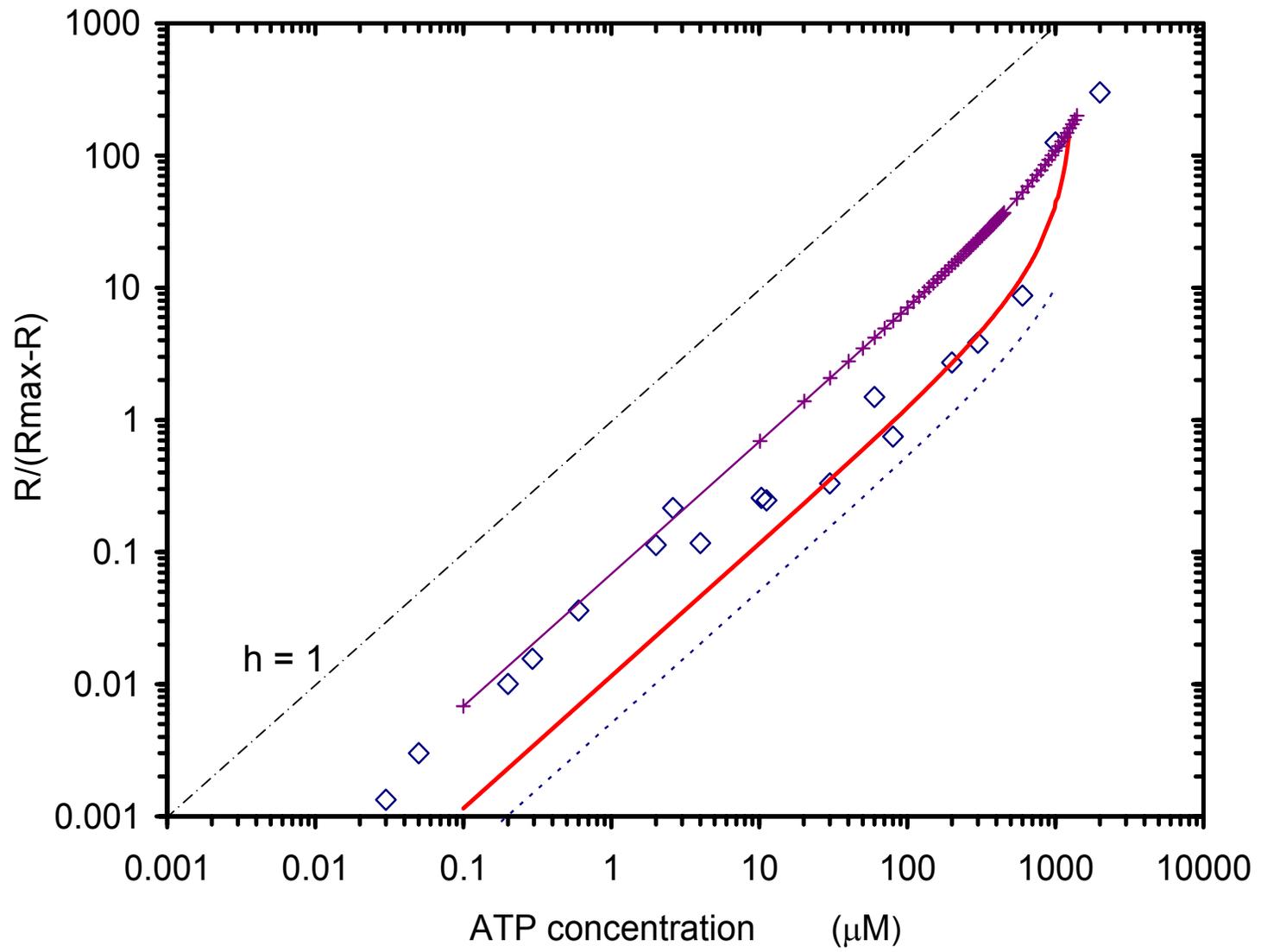

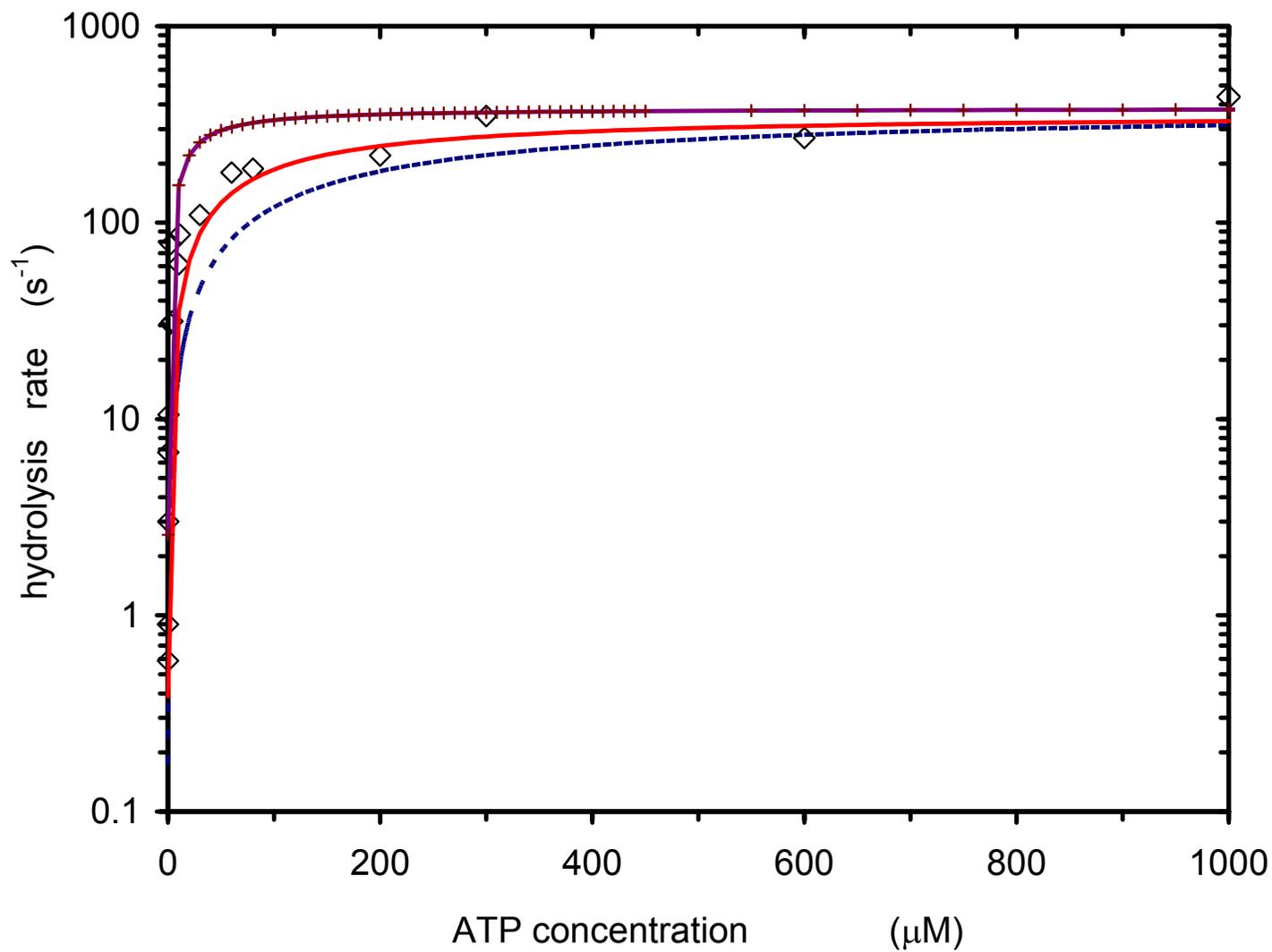

Liu et al., Figure 4

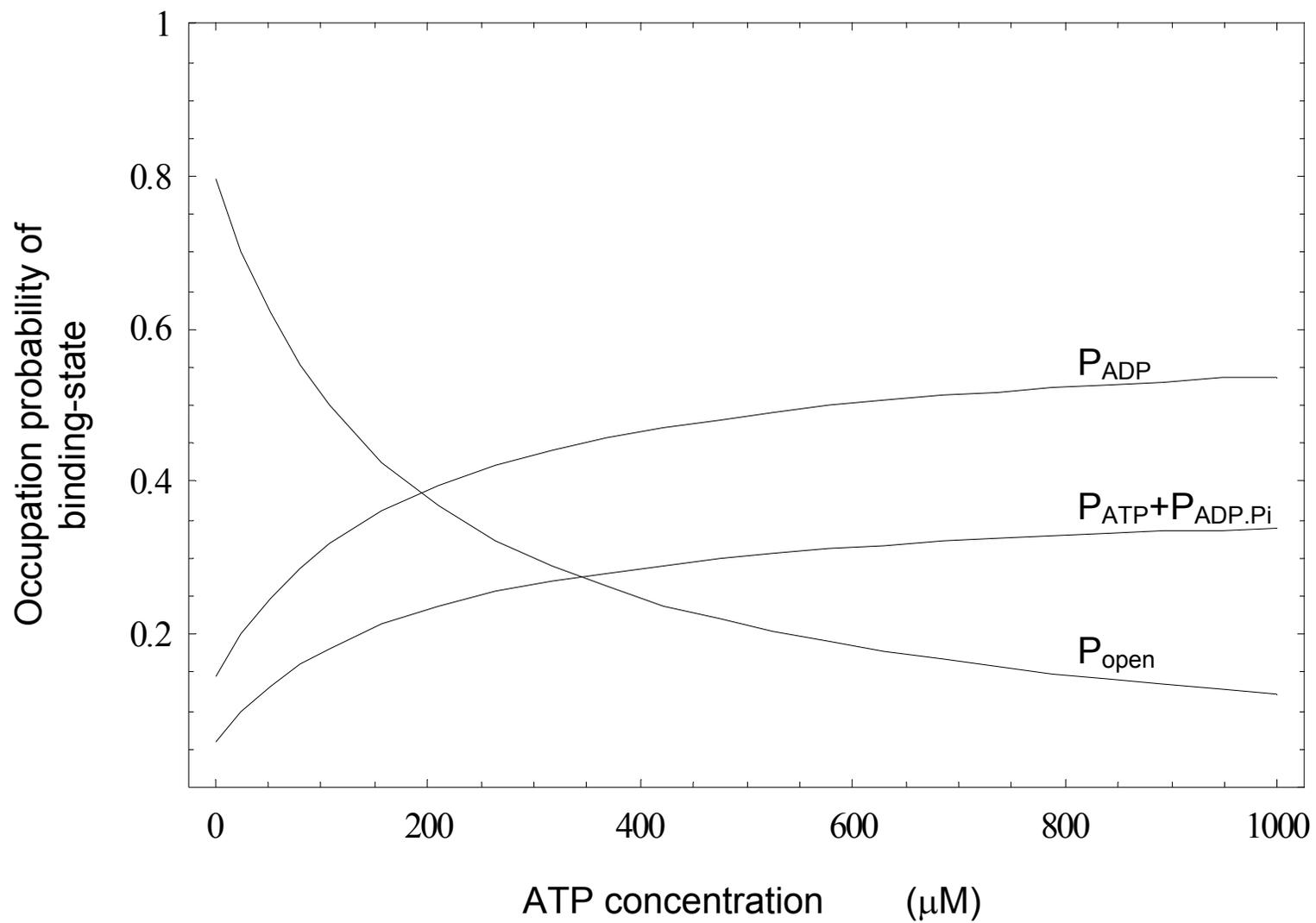